\documentclass[%
aps,
rsi,%
 amsmath,amssymb,
 reprint,%
floatfix]{revtex4-1}
\usepackage{hyperref}
\usepackage{graphicx}
\usepackage{dcolumn}
\usepackage{bm}
\usepackage[latin1,utf8]{inputenc}
\usepackage{float}
\usepackage{xcolor}
\usepackage{tikz}
\usepackage{comment}
\usepackage[toc]{appendix}
\hypersetup{
     colorlinks = true,
     linkcolor = blue,
     anchorcolor = blue,
     citecolor = blue,
     filecolor = blue,
     urlcolor = blue
     }
     
\usepackage{physics}
\usepackage{bbold}
\usepackage{cancel}

\newcommand{\el}[1]{{\color{black} #1}}
 
\begin{document}

\preprint{AIP/123-QED}

\title{Strong-damping limit of quantum Brownian motion in a disordered environment}

\author{Arthur M. Faria$^{1,2}$}
\author{Marcus V. S. Bonança$^1$}
\author{Eric Lutz$^2$}
\affiliation{$^1$Instituto de Física Gleb Wataghin, Universidade Estadual de Campinas,   13083-859 Campinas, São Paulo, Brazil
\\
$^2$
Institute for Theoretical Physics I, University of Stuttgart, D-70550 Stuttgart, Germany
}%


\begin{abstract} We consider a microscopic model of an inhomogeneous environment where an arbitrary quantum system is locally coupled to a harmonic bath via a finite-range interaction. We  show that  in the overdamped regime the position distribution obeys a classical Kramers-Moyal  equation that involves an infinite number of higher derivatives, implying that the  finite bath correlation length  leads to non-Gaussian Markovian noise. We analytically solve the equation for a harmonically bound particle and analyze its non-Gaussian diffusion  as well  as its steady-state properties.
\end{abstract}

\maketitle


\section{\label{sec:level1}Introduction}

The description of quantum systems coupled to an external environment is  of central importance in many different fields, including condensed matter physics \cite{wei12}, quantum optics \cite{scu10} and quantum information science \cite{nie02}, as well as the theory of open quantum systems \cite{bre07}, quantum dissipation \cite{cal14}, and environment-induced decoherence \cite{sch07}. Starting from a microscopic model of the environment, usually a bath of harmonic oscillators \cite{wei12,scu10,cal14,bre07,sch07,nie02}, one  aims at describing the dynamics of the system of interest in terms of approximate master equations for its reduced density operator. In the high-temperature limit, such master equations reduce to the Klein-Kramers equation for phase-space distributions known from the theory of  classical Brownian motion \cite{ris96,cof04}. On the other hand, in the strong system-bath coupling domain, the almost classical dynamics can be described with the help of a Fokker-Planck equation whose noise properties are strongly impacted by quantum fluctuations \cite{ank01,mac04,ank05,luc05,cof07,cof08,dil09,mai10,sin11,def11}. This  so-called quantum Smoluchowski regime has recently been investigated experimentally using superconducting tunnel junctions at ultralow temperatures \cite{jac17}.

The standard theory of quantum Brownian motion assumes that the system is coupled to the bath via a global, infinite-range interaction, where the system-bath coupling is homogeneous in the  position of the system \cite{wei12,scu10,cal14,bre07,sch07,nie02}. There is, however, a large class of microscopic objects where the bath coupling is  local, with a finite interaction range given by a typical length scale of the surroundings. Concrete examples include Brownian motion in  a scattering environment made of a gas of molecules (when the coupling is limited by  the  wavelength of the environment) \cite{gal90,dio95,gal96,vac00,horn06,vac09} and diffusion in a disordered medium (when the coupling is limited by the  spatial correlation length of the disorder potential) \cite{coh97,coh97a,ang97,coh98,bul98,gui03}. The restricted range of the system-bath interaction will in general strongly affect the dynamics of the system, including its dephasing and diffusion properties \cite{gal90,dio95,gal96,vac00,horn06,vac09,coh97,coh97a,ang97,coh98,bul98,gui03}. These studies all consider  weak coupling between system and reservoir.

In this paper, we derive a generalized  Smoluchowski equation for an arbitrary  quantum system strongly coupled to an (ohmic) bath of harmonic oscillators via a finite-range interaction, which can be regarded as a model of a disordered environment (Sec.~II)  \cite{coh97,coh97a,ang97,coh98}. We obtain a classical Kramers-Moyal  equation that involves an infinite number of higher derivatives, and thus includes both Gaussian and non-Gaussian white noise contributions (Sec.~III) \cite{gar97,kam07,rei16}. The usual Schmoluchowski equation with purely Gaussian noise, which corresponds to the first two terms of the expansion, is recovered in the limit of an infinite interaction range. Finally, we analytically determine the  solution of the generalized Smoluchowski equation for the case of a  Brownian particle trapped in a harmonic potential. We  analyze  its non-Gaussian properties as well as  its steady-state distribution (Sec.~IV), thus providing useful insight into the effect of a finite bath correlation length.


\section{\label{sec:level2}Model for local bath coupling}
We consider a general one-dimensional quantum system, with position $x$ and momentum $p$, that is locally coupled to a heat bath made of infinitely many harmonic oscillators,  with canonical coordinates $\left(q_n,p_n\right)$, mass $m_n$  and frequency $\omega_n$ (Fig.~1).
 The total Hamiltonian reads
    \begin{align}
	        \mathcal{H}=\mathcal{H}_{S}+\mathcal{H}_{I}+\mathcal{H}_{B},
	        \label{eq:hamilt}
    \end{align}
   where   the respective system, interaction and bath Hamiltonians, $\mathcal{H}_{S}$,  $\mathcal{H}_{I}$ and  $\mathcal{H}_{B}$ are given by \cite{coh97,coh97a,ang97,coh98}
    \begin{align}
            \label{eq:HS}
	        &\mathcal{H}_{S} = \frac{p^2}{2} + V(x),   
	        \\
	        \label{eq:HI}
	        &\mathcal{H}_{I} = -\sum_{n}c_n \,q_n u(x-x_n), 
	        \\
	        &\mathcal{H}_{B} = \sum_{n}\frac{p_{n}^2}{2m_n} + \frac{1}{2}m_n \omega_n^2 q_n^2 .
	        \label{eq:HB}
    \end{align}
     The system, with mass $m=1$, moves in a confining potential $V(x)$. The scattering potential  $u(x-x_n)$ further specifies  the local interaction between the system and the bath; it depends solely on the distance between the system and the fixed location $x_n$ of $n$-th bath oscillator. The disordered environment is characterized by the spatial autocorrelation function   $w(r) =  \int_{-\infty}^{\infty}dx\, u(r-x)u(x)$, with a finite correlation length $l$. For concreteness, we will use the Gaussian form,
    $w(r) = l^2 \, \exp\left[-r^2/2l^2\right]$, in the following \cite{coh97,coh97a,ang97,coh98}. The spectral density of the bath, $J(\omega) = (\pi/2) \sum_n(c_n^2/m_n\omega_n) \delta (\omega-\omega_n)\sim \omega$ is additionally assumed to be linear at low frequencies, corresponding to an Ohmic, that is, Markovian bath ($c_n$ are the bath coupling constants) \cite{wei12,scu10,cal14,bre07,sch07,nie02}. Hamiltonian (1)  is often referred to as 'diffusion, localization and dissipation' model \cite{coh97,coh97a} or 'mattress' model \cite{ang97}.
    
     \begin{figure}[t!]
        \centering       
        \begin{tikzpicture}
            \node (a) [label={[label distance=-.35 cm]140:\textbf{a)}}]  at (-8.2,0.) {\includegraphics[width=0.17\textwidth]{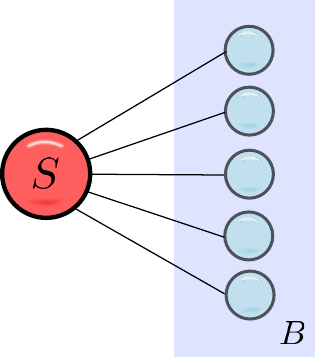}};
            \node (a) [label={[label distance=-.35 cm]140:\textbf{b)}}]  at (-4,0) {\includegraphics[width=0.17 \textwidth]{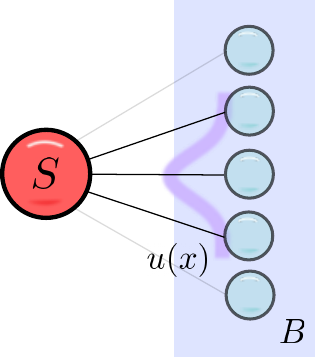}};
	\end{tikzpicture} 
        \caption{Schematic representation of the system-bath interaction. a) A quantum system $S$ is globally coupled to a bath $B$ of harmonic oscillators via an infinite-range interaction. b) A local system-bath coupling   with a finite spatial correlation length is  characterized by a finite-range  interaction $u(x)$.}
        \label{fig:2}
    \end{figure}

    In the high-temperature limit, the reduced Wigner function $W(x,p)$ of the system satisfies the Markovian phase-space equation \cite{coh98}
    \begin{align}
        \nonumber \partial_t W &= -p\,\partial_xW +V'(x)\partial_pW
        \\
        &\hspace{4mm}+  \gamma \partial_p \left[G_F *(p\,W)\right] +\nu \left( G_N *W\right),
        \label{eq:DLD_kramers}
    \end{align}
    where $\gamma$ and $ \nu$ denote  the respective friction and diffusion coefficients.  The two functions $G_F(p)$ and $G_N(p)$ are the corresponding friction and noise kernels, whose expressions  are given in terms of Fourier transforms (denoted by $\mathcal{FT}$) of the correlation function $w(r)$   as \cite{coh98}
        \begin{align}
       \!\! G_F(p) = \mathcal{FT}_{r \to p/\hslash}\left\{-\frac{w'(r)}{r}\right\}
        = \frac{1}{\sqrt{2\pi}(\hslash/l)}e^{-\frac{(p-p')^2}{2(\hslash/l)^2}},\!\!
        \label{eq:GFP}
    \end{align}
    and
     \begin{align}
        \nonumber G_N(p) &= \frac{1}{\hslash^2}\mathcal{FT}_{r \to p/\hslash} \left\{\left[w(r) - w(0)\right]\right\}
        \\
        &= \frac{1}{(\hslash/l)^2}\left[\frac{1}{\sqrt{2\pi}(\hslash/l)} e^{-\frac{(p-p')^2}{2(\hslash/l)^2}}- \delta\left(p-p'\right)\right].
        \label{eq:GNP}
    \end{align}
    Both expressions, that are nonlocal in $p$,  capture  the effects of a finite bath correlation length $l$. The symbol $*$ represents the convolution operation.
    
    In the limit of a global system-bath interaction, $l \to \infty$ (no disorder), $w(r) \to -r^2/2$. As a result, $G_F(p)$ tends to $\delta(p-p')$ and the operator $G_N(p) * (\cdot)$ can be replaced by $\partial_p^2$ \cite{coh98}. In this case, Eq.~\eqref{eq:DLD_kramers}
 reduces to the familiar Klein-Kramers equation for classical Brownian motion \cite{ris96,cof04}
    \begin{align}
        \partial_t W = -p\,\partial_xW +V'(x)\partial_pW + \gamma \partial_p \left[pW\right] + \nu\partial_p^2W.
    \end{align}


\section{Strong-damping limit \label{sec:StrongDamp}}

 We next consider the strong-friction limit of Eq.~\eqref{eq:DLD_kramers}. For large damping, the momentum  is expected to quickly relax to its stationary limit, so that (for infinite friction), the phase-space distribution can be factorized as $ W(x,p,t) = \varrho_{s}(p) \rho(x,t)$, where 
    $\varrho_{s}(p)$ denotes the stationary momentum distribution \cite{ris96,cof04}. In the following, we shall derive a generalized Smoluchowski equation by  eliminating the fast momentum variable \cite{kra40,cha43}. 
    
  We  begin by rewriting the phase-space equation \eqref{eq:DLD_kramers} in the form of an infinite sum that will be more convenient to analyze. To that end, we note that the friction and noise kernels can be expressed as (Appendix A)
        \begin{align}
        \label{eq:gf}
        &G_F(p) * \bm{[}\,p\,\bm{\cdot\,]} \to  \exp(\frac{(\hslash/l)^2\partial_p^2}{2}) \bm{[}\,p\,\bm{\cdot\,]}, 
        \\
        &G_N(p) * \bm{[\,\cdot\,]} \to  \frac{1}{(\hslash/l)^2}\left(\exp(\frac{(\hslash/l)^2\partial_p^2}{2}) - 1\right) \bm{[\,\cdot\,]}.
        \label{eq:kernel_exp}
    \end{align}
    By expanding the two exponentials in Eqs.~\eqref{eq:gf} and \eqref{eq:kernel_exp}
 in a power series, we obtain
        \begin{align}
        \partial_t W &=-p\partial_xW+ V'(x) \partial_p W + \gamma \partial_p\left[p\mathcal{W}\right] + \nu\partial_p^2W
       \nonumber \\
        &  + \gamma\sum_{n=1}^{\infty} \gamma_{n}\partial_p^{2n+1}\left[p W\right]
        + \frac{\nu}{(\hslash/l)^2}\sum_{n=2}^{\infty} \gamma_{n} \partial_p^{2n} W.
        \label{eq:ME_nonG1}
    \end{align}
     where $\gamma_n = (\hslash/l)^{2n}/(2^{n} n!)$ (for simplicity, we  will henceforth set \(\hslash = 1\)). The first line corresponds to the standard Klein-Kramers equation ($l \to \infty$), while the two infinite sums on the second line, that involve all higher derivatives, describe the influence of a finite length  $l$.
    
  \begin{figure*}[t!]
	\centering
	\begin{tikzpicture}
            \node (a) [label={[label distance=-.35 cm]145:\textbf{a)}}]  at (-8.2,-0.08) {\includegraphics[width=0.435\textwidth]{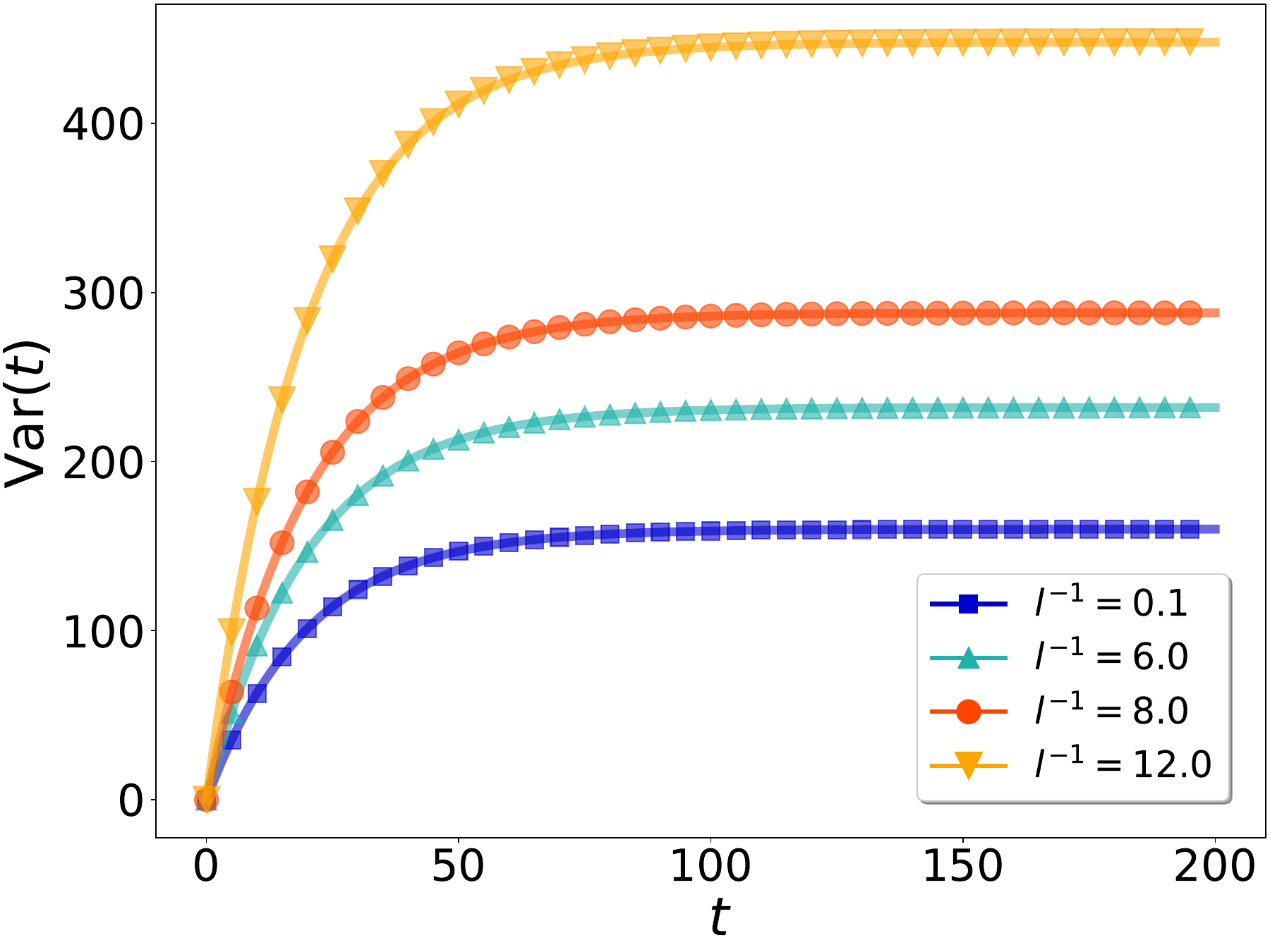}};
            \node (a) [label={[label distance=-.35 cm]145:\textbf{b)}}]  at (1.,0) {\includegraphics[width=0.44 \textwidth]{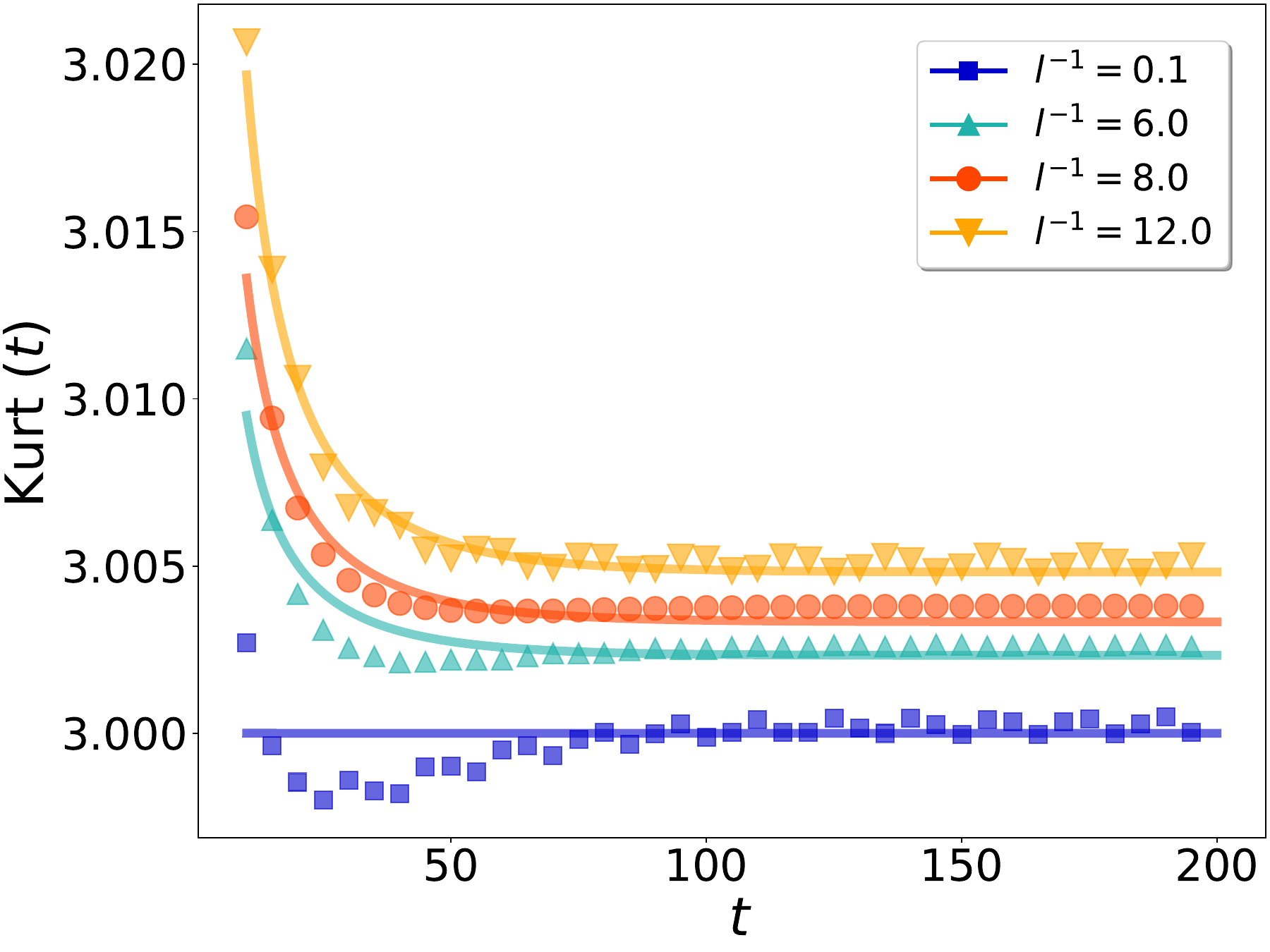}};
	\end{tikzpicture}
        \caption{a) Time evolution of the position variance $\text{Var}(t) $ for a harmonically bound particle for five values of the inverse bath correlation length, $l^{-1} = 0.1, 6.0, 8.0, 12.0$. Symbols correspond to numerical solutions of the generalized Smoluchowski equation, \eqref{eq:DLD_smol1}, up to fifth order, while solid lines correspond to the analytical expression \eqref{eq:variance_t}.  A finite  value of the correlation length $l$ enhances the diffusion. b) Corresponding kurtosis $\text{Kur}(t) $, where the solid lines correspond to the analytical expression \eqref{eq:kurtosis_t}. A finite value of $l$ increases the non-Gaussianity of the process.  Parameters are $kT = 40.0$, $\gamma = 10.0$ and \mbox{$\kappa = 0.5$, \el{yielding $\varepsilon = 80$}}.}
        \label{fig:moments}
    \end{figure*}

Following Refs.~\cite{kra40,cha43} (see also Ref.~\cite{cof04}), we  rewrite Eq.~\eqref{eq:ME_nonG1} in the form
      \begin{align}
        \nonumber\partial_t W   &= \frac{1}{\gamma}\partial_x\left[V'(x)W\right] +\nu\partial_p^{2}W
        \\
        \nonumber & \hspace{3mm}+\gamma\left(\partial_p - \frac{1}{\gamma}\partial_x \right)\left(p W  +\frac{ V'(x)}{\gamma} W \right) 
        \\
        & \hspace{3mm}+  \gamma\sum_{n=1}^{\infty} \gamma_n
        \partial_p^{2n+1} \left[pW\right] + \nu l^2 \sum_{n=2}^{\infty}\gamma_{n}\partial_p^{2n}W,
        \label{eq:chad_method}
    \end{align}
  and proceed by integrating both sides of the equation  along the straight phase-space line $x+ \gamma^{-1}p = x'$ with line element $ds$
  \cite{cof04}. In the strong damping limit,  the marginal $\rho(x_,t)$ follows from $W(x,p)$ as
    \begin{align}
\int_{x+ \gamma^{-1}p }ds\, W \ \underset{\gamma \to \infty}{\to} \ \int dp\,  W  = \rho(x',t).
    \end{align}
    In addition, we have  $\partial_{x}^n = \partial_{x'}^n$ and $\partial_p^n = (-\gamma)^{-n}\partial_{x}^n$ along the integration path. As a consequence,  $\partial_p - \gamma^{-1}\partial_{x'}$ is a null operator along this path, which  eliminates the contribution coming from the first two terms in Eq.~\eqref{eq:chad_method}. Moreover, since $\int ds\,\partial_p^{2n+1} [p\,W] \to_{\gamma \to \infty}  \gamma^{-2n}\,\partial_{x'}^{2n} \rho$,  the fourth term in Eq.~\eqref{eq:chad_method} reduces to
    \begin{align}
           \!\!  \gamma\sum_{n=1}^{\infty} \gamma_n\int ds\,\partial_p^{2n+1} [p\,W]  \underset{\gamma \to \infty}{\to} \gamma\sum_{n=1}^{\infty}\frac{(1/\gamma l)^{2n}}{2^n n!} \, \partial_{x'}^{2n} \rho.\!\!
    \end{align}
 A similar  calculation applied to  the fifth term yields
     \begin{align}
       \nu l^2 \sum_{n=2}^{\infty}\gamma_{n}\int ds\, \partial_p^{2n}W \underset{\gamma \to \infty}{\to} \nu l^2\sum_{n=2}^{\infty} \frac{\gamma_{n}}{\gamma^{2n}}  \partial_{x'}^{2n}\rho.
    \end{align}
   Combining all the terms, we obtain the generalized  overdamped Smoluchowski equation
     \begin{align}
        \partial_t \rho  &= \frac{1}{\gamma}\partial_x\left[V'(x)\rho\right]  + \sum_{n=1}^{\infty} \frac{a_{2n}}{(2n)!} \partial_{x}^{2n} \rho,
        \label{eq:DLD_smol1}
    \end{align}
    where $ a_{2n}= (\gamma  + {\nu l^2})(2n)!/[{2^{n} }{(\gamma l)^{2n}n!}]$; without loss of generality, we have replaced $x' \to x$. Equation \eqref{eq:DLD_smol1} has the form of a classical Kramers-Moyal equation that includes both Gaussian and non-Gaussian white noise contributions \cite{gar97,kam07}: The first two terms  describe   deterministic drift plus continuous Gaussian diffusion, whereas the higher-order terms ($n\geq 2$) correspond to discrete jumps and to non-Gaussian white noise \cite{gar97,kam07}. In the limit of a global system-bath interaction, $l \to \infty$, 
     the standard overdamped Smoluchowski equation is recovered
    \begin{align}
        \partial_t \rho  &= \frac{1}{\gamma}\partial_x\left[V'(x)\rho\right] +  \frac{\nu}{2\gamma^2}  \partial_{x}^{2}\rho\,,
    \end{align}
    with only Gaussian noise.
Note that in this case the  Einstein relation, $ \nu /2\gamma  =kT $ \cite{gar97,kam07}, where $k$ the Boltzmann constant,  implies the coefficients
    \begin{align}
        a_{2n} = \frac{\gamma \left[1  + 2kTl^2\right]
}{2^{n} n! (\gamma l)^{2n}}   .     \label{eq:Dn}
    \end{align}
  We may thus   conclude that the effect of a finite correlation length $l$ is  the occurrence of discrete non-Gaussian Markovian jump events in the overdamped dynamics. It is worthwhile mentioning that this strongly differs from  the coupling to a finite number of bath modes, which usually leads to  non-Markovian dynamics \cite{riv14,bre16,veg17}. In the present case, the system is still coupled to all the bath modes, but with a modulated coupling strength. 
     \begin{figure*}[!t]
	\centering
	\begin{tikzpicture}
        \node (a) [label={[label distance=-.35 cm]145:\textbf{a)}}]  at (-8.2,-0.08) {\includegraphics[width=0.435\textwidth]{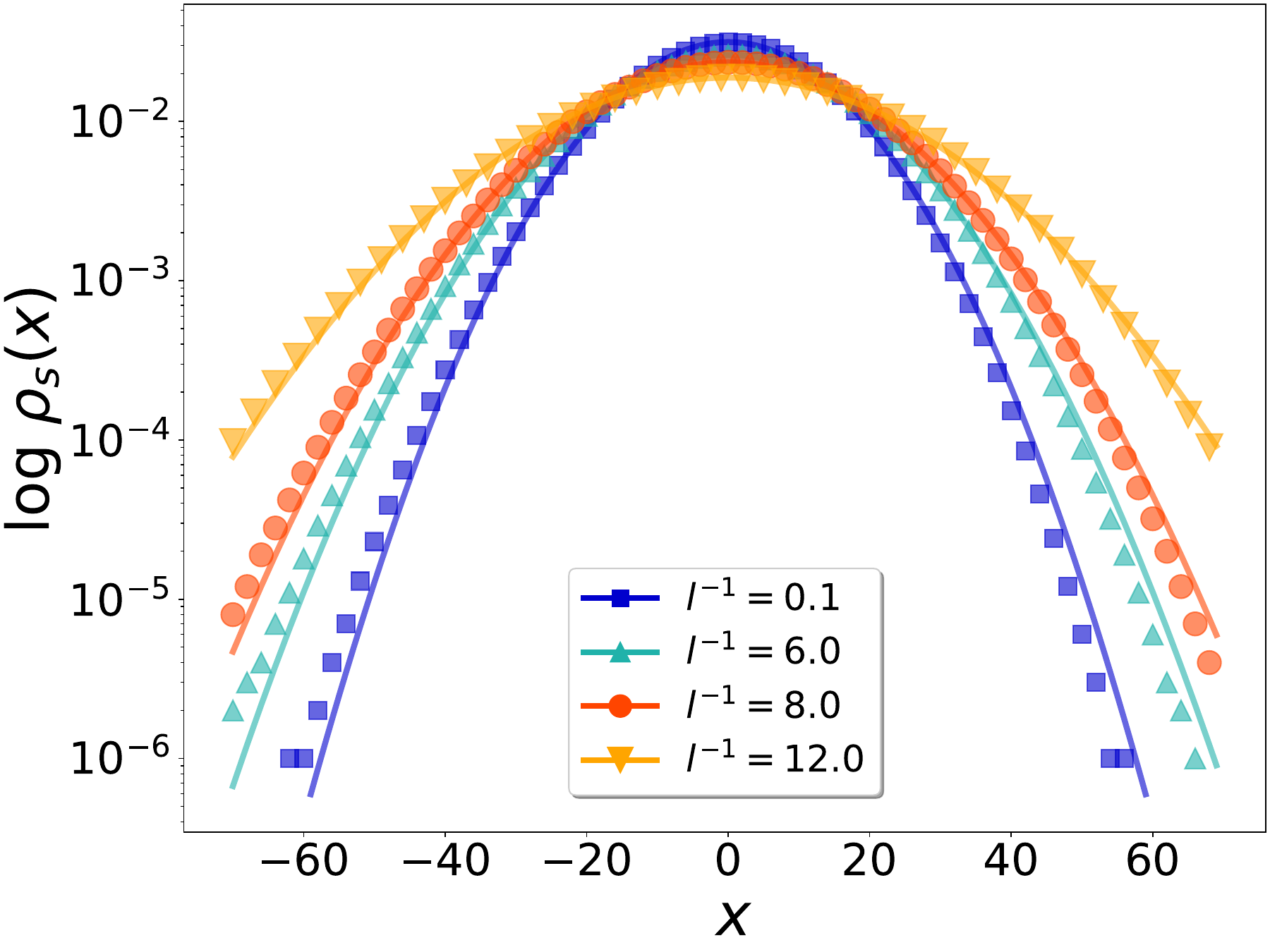}};
        \node (a) [label={[label distance=-.35 cm]145:\textbf{b)}}]  at (1.,0) {\includegraphics[width=0.44 \textwidth]{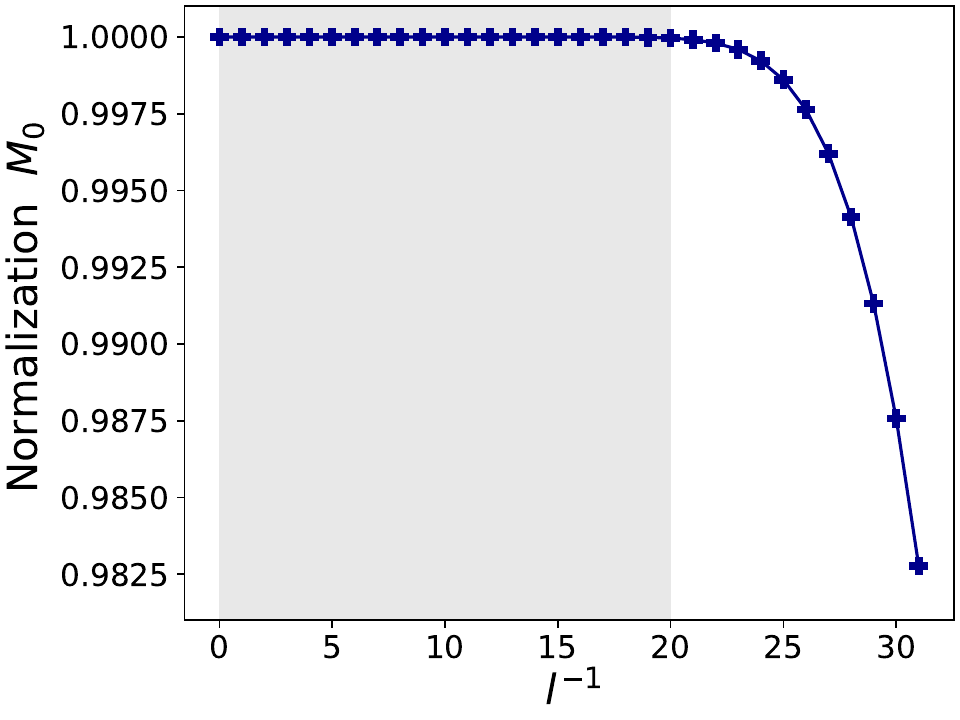}};
	\end{tikzpicture}
        \caption{a) Steady-state position distribution $\rho_s(x)$ for a harmonically bound particle for five values of the inverse bath correlation length, $l^{-1} = 0.1, 6.0, 8.0, 12.0$. Symbols correspond to numerical solutions of the generalized Smoluchowski equation \eqref{eq:DLD_smol1}, up to fifth order, while solid lines correspond to the analytical saddle-point approximation \eqref{eq:steady state}. Non-Gaussian tails appear as $l^{-1}$ departs from the Gaussian limit $l^{-1}=0$. b) The steady-state distribution is only properly normalized, with unit zeroth moment $M_0$, for small values of $l^{-1}$. The shaded gray area represents the analytical estimate $l^{-1} < l^{-1}_{min}$ where $l^{-1}_{min} =2 \gamma$ is determined by the damping coefficient $\gamma$.  Parameters  are  $kT = 40.0$, $\gamma = 10.0$ and $\kappa = 0.5$, yielding $\varepsilon = 80$.}
        \label{fig:rho_sigma}
    \end{figure*}

    

\section{ Harmonically bound particle \label{sec:SteadySt}}
Let us now investigate the dynamics of a harmonically bound particle with potential $V(x) = \kappa x^2/2$, where  $\kappa$ is the spring constant.
For this purpose, we solve Eq.~\eqref{eq:DLD_smol1} in Fourier space, and introduce the characteristic function
$\varphi(z,t) = \mathcal{FT}_{x\to z}\left\{ \rho(x,t)\right\}$ which is defined via \cite{ris96,cof04}
     \begin{align}
        \rho(x,t) = \frac{1}{2\pi}\int_{-\infty}^{\infty}dz\, e^{izx}\varphi(z,t).
        \label{eq:rho_IFT}
    \end{align}
The  master equation for $\varphi(z,t)$ accordingly reads
      \begin{align}
        \frac{\partial \varphi}{\partial t}  &=  -\frac{\kappa }{\gamma} z \frac{\partial \varphi}{\partial z}  + \frac{kT}{\gamma}\sum_{n=1}^{\infty} D_{2n}  (iz)^{2n}\varphi,
        \label{eq:master_phi}
    \end{align}
    with the coefficients $D_{2n} = \gamma a_{2n}/kT (2n)!$. Equation \eqref{eq:master_phi} can be solved with the  method of characteristics \cite{ris96,cof04}. With the initial condition $\rho(x,0) = \delta(x)$, we find
        \begin{align}
        \varphi(z,t)  
                \nonumber &= \exp\left[\varepsilon \frac{D_{2} }{2}(iz)^2(1 - e^{-2 \kappa t/\gamma})\right]
        \\
        \nonumber&\times\exp\left[\varepsilon \sum_{n=2}^{\infty} \frac{D_{2n} }{2n}(iz)^{2n}\left(1 - e^{-2 n\kappa t/\gamma}\right)\right]\\
        &= \varphi^{G}(z,t)\varphi^{NG}(z,t) ,
        \label{eq:charac}
    \end{align}
     where  $\varepsilon = kT/\kappa$. We observe that  $\varphi(z,t)$ can be written as a product of  two, Gaussian and  non-Gaussian, components, $  \varphi^{G}(z,t)$ and $\varphi^{NG}(z,t)$, where the latter term encodes the  contribution of the finite bath correlation length $l$. The asymptotic steady states $\varphi_s(z)$ is moreover approached exponentially in time.

         Additional insight into the time evolution of the system and the effect of the inhomogeneous bath coupling  may be gained by evaluating the moments of the position distribution $\rho(x)$, $M_{m}(t)= i^{m}\left.\partial_z^{m} \varphi(z,t)\right|_{z=0}$ \cite{ris96,cof04}.   Variance    and kurtosis are explicitly given by
   \begin{align}
               \label{eq:variance_t}
        &\text{Var}(t) = M_2(t) = \frac{1+ 2 kT l^{2}}{2\kappa l^2 }(1 - e^{-2 t \kappa/\gamma}),
        \\
        &\text{Kurt}(t) = \frac{M_4(t)}{\text{Var}(t)^2} = 3 +  \frac{3  \kappa \coth{\left(t\kappa/\gamma\right) }}{\gamma^2\left(1+ 2 kTl^2\right)} .
        \label{eq:kurtosis_t}
    \end{align} 
    We note that all the odd moments ($m\geq3$) of $\rho(x)$ vanish, since the Kramers-Moyal expansion \eqref{eq:DLD_smol1} only contains even higher derivatives ($n\geq 2$). We additionally mention that the exponential relaxation of the mean, $M_1(t)=x_0e^{- \kappa t/\gamma}$,  from an initial position $x_0\neq0$ is not affected by the value of $l$. For long times, we further have
        \begin{align}
        \label{eq:variance}
        &\text{Var} =\lim_{t\to \infty}\text{Var}(t)  = \frac{1+ 2 kT l^{2}}{2\kappa l^2 }\,,
        \\  \label{eq:kurtosis}
              &\text{Kurt} =\lim_{t\to \infty}\text{Kurt}(t) = 3 +  \frac{3\kappa}{\gamma^2\left(1+ 2 kTl^{2}\right)}\,.
    \end{align}
   A finite value of the bath correlation length $l$ therefore enhances the value of the variance above its Gaussian limit (Fig.~2a), indicating that the system diffuses faster owing to the additional non-Gaussian fluctuations. \el{At} the same time, it increases the kurtosis, making the position distribution non-Gaussian (Fig.~2b).

  The case of free diffusion can be easily recovered by taking the limit  $\kappa \to 0$. The corresponding variance reads     
    \begin{align}
        \text{Var}_0(t) = \left(\frac{1 + 2kTl^{2}}{\kappa l^2}\right)t,
        \label{eq:free_dif}
    \end{align}
showing that it is linear in time, as for normal Brownian motion \cite{ris96,cof04}, albeit with an enhanced diffusion constant, $D=(1 + 2kTl^{2})/(2\kappa l^2)$, which can be regarded as a generalized Einstein relation \cite{ris96,cof04}. We hence have the remarkable result that the local bath coupling does not modify the linear time-dependence of the variance. The diffusion constant of standard Brownian motion, $D_0=kT/\kappa$ is again obtained for  an infinite correlation length.

    
   We proceed by determining the steady-state distribution $\rho_s(x)$ which is given by the inverse Fourier transform
       \begin{align}
        \rho_s(x) = \frac{1}{2\pi}\int_{-\infty}^{\infty}dz\, e^{\varepsilon \Gamma(z)},
        \label{eq:rho_IFT_saddle}
    \end{align}
    where the exponent reads
    \begin{align}
        \Gamma(z) &= \frac{iz}{\varepsilon} x + \varphi_s(z) = \frac{iz}{\varepsilon} x + \sum_{n=1}^{\infty} \frac{D_{2n} }{2n} (iz)^{2n}.
        \label{eq:G}
    \end{align}
     The integral \eqref{eq:rho_IFT_saddle} cannot be evaluated exactly. However, since $\varepsilon = kT /\kappa\gg 1$ a saddle-point method is applicable \cite{ben13}. In the limit $\varepsilon \gg 1$, the integral is indeed dominated by the maximum, $z=q$,  of the function $\Gamma(z)$ which satisfies $\partial_z \Gamma = 0$. To leading order, one has \cite{ben13}
     \begin{align}
     \rho_{s}(x) = \exp[{\varepsilon\Gamma(q)}]/\sqrt{2\pi\varepsilon\left|\partial_z^2\Gamma(q)\right|}.
     \end{align}
   Setting the derivative of Eq.~\eqref{eq:G} to zero, we find
    \begin{align}
        \partial_q\varphi(q)  &=\sum_{n=1}^{\infty}  (-1)^nD_{2n} (q)^{2n-1} = - \frac{ix}{\varepsilon} ,
        \label{eq:first_der}
    \end{align}
 whose solution determines $q=q(x)$. We further have 
   \begin{align}
        \left|\partial_q^2\Gamma\right| &= \left|\sum_{n=1}^{\infty} (2n-1)D_{2n} \, (iq)^{2n-2}\right|.
        \label{eq:second_der}
    \end{align}
    In order to solve Eq.~\eqref{eq:first_der}, we note that since $\varepsilon$ is large, we may, to leading order, use the approximation $q = q_0 + \mathcal{O}(1/\varepsilon)$. We hence obtain
    \begin{align}
        iq_0 = -\frac{x}{\varepsilon D_{2}} = -\frac{2 \kappa x}{\gamma a_2}.
        \label{eq:iq}
    \end{align}
    Combining everything, we eventually arrive at the steady-state position distribution  
        \begin{align}
           \rho_{s}(x)   \nonumber   &= \frac{\sqrt{2\pi\gamma/\kappa}}{\left|\sum_{n=1}^{\infty} \frac{2n-1}{(2n)!}a_{2n} \, \left(\frac{2\kappa x}{\gamma a_{2}} \right)^{2n-2}\right|^{1/2}}\\ &\times \exp\hspace{-0.3mm}\left[-\frac{\kappa x^2}{\gamma a_{2}}\right]       \exp\hspace{-0.3mm}\left[ \frac{\gamma}{\kappa}\sum_{n=2}^{\infty} \hspace{-0.5mm}\frac{a_{2n}}{2n(2n)!} \hspace{-0.5mm}\left(\frac{2\kappa x}{\gamma a_{2}}\right)^{2n}\hspace{-0.5mm}\right]\nonumber \\
           &= \rho_{s}^{G}(x)\rho_{s}^{NG}(x).
            \label{eq:steady state}
        \end{align}
    As before, the stationary distribution can be expressed as a product of Gaussian and non-Gaussian contributions, $\rho_{s}^{G}(x)$ and $\rho_{s}^{NG}(x)$. Since $a_{2} \to 2kT/\gamma$ and $a_{2n} \to 0$ (for $n>2$),  when $l \to \infty$,  the exact Gaussian position distribution is recovered in the limit of large $l$ \cite{ris96,cof04}
     \begin{align}
        \rho_{s}^{G}(x) &=  \sqrt{\frac{\kappa}{2\pi kT}}  \exp\left[-\frac{\kappa x^2}{2kT} \right].
        \label{eq:st_brow}
    \end{align}
    Figure 3a displays the stationary position distribution $\rho_s(x)$ (i) determined  using the  Kramers-Moyal equation  \eqref{eq:DLD_smol1} by keeping the first five terms (symbols) and ii) given by the saddle-point approximation \eqref{eq:steady state} (solid lines), for different values of the inverse correlation length $l^{-1}$. We observe that the distribution becomes broader, and more and more non-Gaussian, as $l$ decreases, as expected from the previous analysis of the moments  \eqref{eq:variance} and \eqref{eq:kurtosis}. We moreover have excellent agreement between exact and  approximated steady-state solutions.
    
    It is important to realize that due to the infinite sum appearing in the exponent of the non-Gaussian part, the  steady-state solution \eqref{eq:steady state} is not always properly normalized, that is,   the zeroth moment of the distribution, $M_0$,  is not always equal to one. This is the case for  small values of the  correlation length $l$, when the  distribution becomes too broad. 
     We estimate the minimal value of the parameter $l$ that leads to a properly normalizable stationary position distribution, by analyzing the convergence of the infinite sum that appears in  the Kramers-Moyal equation \eqref{eq:DLD_smol1}. A necessary condition for convergence is that  $a_{2(n+1)}/(2(n+1))!<a_{2n}/(2n)!$ \cite{ril06}. In particular, the condition $a_4/4< a_2/2$  implies that $l>l_{min} = 1/2\gamma$. The minimal value of the correlation length is therefore  determined by the damping coefficient $\gamma$. Figure 3b shows the normalization $M_0$ of the stationary position distribution \eqref{eq:steady state} as a function of $l^{-1}$ for $\gamma=10$. The zeroth moment is equal to one until the value $l^{-1} = 20$ is reached, after which it sharply drops. This threshold agrees with the estimate  $ l_{min}^{-1} = 2\gamma= 20$ (gray shaded area). We  further note that the condition $a_{2n}\ll a_2$ ensures a positive  distribution $\rho(x,t)$ when the Kramers-Moyal equation is truncated at a finite order ($n>2$) \cite{ris96,cof04}.


\section{Conclusions \label{sec:Conclu}}

The microscopic origin of non-Gaussian noise in systems coupled to a heat bath is still unclear \cite{kan15,kan15a,kan17}. Arguments based on the central-limit theorem indeed  always predict Gaussian noise \cite{kam07,rei16}. In the present study, we have found that non-Gaussian noise naturally occurs in the overdamped limit of a quantum particle locally coupled to an inhomogeneous high-temperature environment that can be modeled as an ensemble of linear harmonic oscillators. Contrary to the usual case of infinite-range bath coupling,  a finite spatial correlation length leads to a position distribution that evolves according to a Kramers-Moyal equation involving all higher derivatives. Such Markovian equation describes Gaussian diffusion plus non-Gaussian  discrete jump events. As a consequence, diffusion is boosted by the presence of non-Gaussian fluctuations and the steady-state solution of a damped harmonic oscillator exhibit non-Gaussian tails.


\vspace{5mm}
\begin{acknowledgments}
    \vspace{-0.2cm}
    The authors  acknowledge  financial support from the Brazilian agency CNPq (Project Nos. 142556/2018-1 and 304120/2022-7), FAPESP (Project No 2020/02170-4), as well as from the German Science Foundation DFG (Project No. FOR 2724).
\end{acknowledgments}


\appendix

\begin{widetext}

\section{Expansion of noise and friction kernels}
\label{app:B}

    In this Appendix, we demonstrate the identities given in Eqs.~\eqref{eq:gf}-\eqref{eq:kernel_exp} for the friction and noise operators, \( G_F(p) * (\cdot) \) and \( G_N(p) * (\cdot) \).     
    Applying the Fourier transform (FT) and its inverse (IFT) to the kernels \eqref{eq:GFP} and \eqref{eq:GNP}, we have

    \begin{align}
        G_F(p) * [p\, W] &= \mathcal{IFT}_{k\to x}\left\{\widetilde{G}_F(k)\, \mathcal{FT}_{x\to k}\left\{p\, W\right\}\right\} = \frac{1}{\sqrt{2\pi}}\sum_{n=0}^{\infty} \frac{l^{-2n}}{2^{n}n! }\left[\mathcal{IFT}_{k\to x}\left\{ (iz)^{2n}\right\}  * p\, W\right] 
        \\
        &= \sum_{n=0}^{\infty} \frac{l^{-2n}}{2^{n}n! }\left[  \delta^{(2n)} * p\,W\right]     
    =  \sum_{n=0}^{\infty} \frac{l^{-2n}}{2^{n}n! }\frac{\partial ^{2n}}{\partial p^{2n}} [p\,W]  
        = \exp(\frac{l^{-2}\partial_p^2}{2}) [p\,W],
    \end{align}
    \noindent and   
    \begin{align}
        G_N(p) * W &= \mathcal{IFT}_{k\to x}\left\{G_N(k) \varphi(k)\right\} = \frac{1}{\sqrt{2\pi}}\sum_{n=0}^{\infty} \frac{l^{-2n}}{2^{n+1}(n+1)! }\left[\mathcal{IFT}\left\{ (iz)^{2n+ 2}\right\}  * W\right] 
        \\
        &= \sum_{n=0}^{\infty} \frac{l^{-2n}}{2^{n+1}(n+1)! }\left[  \delta^{(2n+ 2)} * W\right] 
        =  \sum_{n=0}^{\infty} \frac{l^{-2n}}{2^{n+1}(n+1)! }\frac{\partial ^{2n+2}}{\partial p^{2n+2}} W 
        =  \frac{1}{l^{-2}}\left(\exp(\frac{l^{-2}\partial_p^2}{2}) - 1\right) [W],
    \end{align}

    \noindent where have used  $\mathcal{IFT}_{k \to x}\left\{ (iz)^{n}\right\} = \sqrt{2\pi}\delta^{(n)}$ and $(\delta^{(n)} * F)(p) =  F^{(n)}(p)$. 

\end{widetext}
   


\end{document}